# The Impact of IT Projects' Complexity on Cost Overruns and Schedule Delays

Mahta Taghi Zadeh
Tecsys Inc., Toronto, Canada.

Rasha Kashef
Ryerson University, Toronto, Ontario, Canada

*Abstract*— This study aims to assist Company's leaders and management team in providing more accurate time and cost estimations for future software projects. This paper focuses on analyzing the relationship between project complexity and cost/time overrun. The sample data of around 50 projects are collected from the Changepoint database. The two research hypotheses were defined and tested using statistical techniques. The quantitative analysis includes descriptive analysis and regression modelling. Experimental results show a strong positive linear relationship between project complexity and cost/time overrun. Based on the study outcome, there is a need to focus on the planning phase of complex projects and allocate more time and budget to estimate the completion date and the initial budget of this type of project more accurately.

## I. INTRODUCTION

"Changepoint is a Business Execution Management™ company, with leading solutions to help organizations better connect corporate strategy with cross-enterprise work planning and management "[1]. Changepoint combines professional services automation (PSA) and project portfolio management (PPM) to provide total visibility and control for business portfolios [1]. It provides professional services software and IT management solutions for corporate businesses and helps executives enhance the effectiveness and productivity of their organization. Changepoint was founded in 1992 in Toronto and was acquired by Compuware Corporation in 2004; the number of employees was around 190 at that time. Changepoint was sold to a Marlin Equity Partners firm on February 2, 2014. It acquired Daptiv Solutions, the leading provider of pure SaaS-based PPM software, on August 21, 2014. This merge provided customers with the most flexible PPM and PSA solutions [2].

Changepoint completes various IT projects with different complexity levels every year. The project's cost and time overruns are the key challenges the company has been faced for a long time. Delays and cost overruns are inevitable in the software development industry since technology is changing continuously. The complexity of projects is defined based on initial duration and budget in Changepoint. This study clarifies any relationship between the complexity of projects and cost/time overruns.

## II. PROBLEM DEFINITION

Changepoint has always suffered from delivery delays or cost overruns of its projects. Cost and time overruns have been the main Company's challenge for a long time. Even though the company adopted various new technologies and tools, it still suffers from these two significant problems, particularly large customers. The level of complexity of projects might be one of the factors contributing to this poor performance, and its role should be more investigated. This study aims to conduct quantitative research to explore the relationships between the complexity of projects and the project cost and time overruns. The outcome of this study will assist the Company's top management in deciding if it would be of benefit to reduce the complexity of projects by breaking down the large-scale and complex projects into smaller pieces and running them as separate standalone projects. This strategy will minimize the risk of time and cost overruns and help the management team have more appropriate supervision on the projects.

## III. RELATED WORK AND LITERATURE REVIEW

A cost overrun occurs when the actual costs or expenses of a project exceed the initial budget. We might also have a delay in completing a project compared to the initial schedule. Several studies have been conducted on cost overruns and time delays in various projects, mostly in the construction industry. The result of a study by Ameh, O.J., and Osegbo [3] shows that the project time overrun is happening all the time and around the globe internationally. "According to Kaming et al. [4] and Trigunarsyah [5], time overrun is the extension of time beyond planned completion dates usually traceable to contractors. Elinwa and Joshua [6] defined it as the time lapse between the agreed estimation or completion date and the actual completion date. Bramble and Callahan [7] describe time overrun as the time during which some part of the construction project is completed beyond the project completion date or not performed as planned due to an unanticipated circumstance. Al-Momani [8] surveyed 130 public projects in Jordan and found delays occurred in 106 projects (82%). Frimpong et al. [9] observed that 33 out of 47 (70%) projects in Ghana were delayed. Ogunlana et al. [10] in Thailand found that most project delays were laid on the contractor. Abd. Majid and McCaffer [11] revealed that 50% of the delays to construction projects could be categorized as non-excusable delays, for which the contractors were responsible. Time overrun affects the project owners, contractors and other project participants. Project owners may be affected through lost benefits that could have accrued from the completed facility. In contrast, contractors may have to spend more on labour and plant, pay penalties as per the contract or even lose other profitable contracts because resources for the next job are tied up on delayed projects". Bertin Bella Akoa from Michigan State University [12] has investigated cost overruns and time delays of construction projects in developing countries using statistical analysis techniques. This study identified causes of cost and time overruns, considering the influence of project size, type, duration and project funding. Chen and Hartman [13] used artificial neural networks to develop a predictive model for project time and cost performance. Their research was an alternative to the multiple linear regression techniques normally used to predict models. Federle and Pigneri [14] used the multiple linear regression methodology to develop a model to predict cost overruns for the Iowa Department of Transportation. This study indicates a relationship between project characteristics, design, the contract award process and cost overruns. A study conducted by the Standish Group on IT projects in 2004 indicates that the majority of IT projects have an average cost overrun of 43 percent, and 71 percent of projects are over budget and time in the US [15]. The 2012 CHAOS results show an increasing trend in the rate of project success. 39% of all projects get delivered on time and budget; 43% had cost and time overruns, and 18% of projects failed completely [16]. Cost overruns estimation has received significant attention, especially in the construction industry [17]-[29]. This paper focuses on quantitative analysis of software development projects as not many quantitative analysis research studies have been found on cost and time overruns in this area.

## IV. Data Collection and Input Analysis

The data for this study has been collected from Changepoint Company. The company uses Cognos business intelligence tool to extract and categorize data. A sample size of fifty projects completed within the last three years has been selected to conduct this research. The data associated with these fifty projects included: overall budget (initial cost), final actual cost, original duration (initial schedule), actual duration and the complexity of projects indicated as five categories listed in Appendix 1. The data associated with fifty projects are listed in Appendix 2. The complexity of projects in Changepoint Company is defined based on the complexity dimension (initial duration and budget).

## V. Quantitative Analysis

The descriptive statistics [30][31][32] including the mean and standard deviation of cost and time overruns for fifty sample projects, were obtained at the first stage. The inferential statistical analysis was then performed to generalize the average of all the company's time and cost overruns. The hypotheses were also defined to determine the correlation between project complexity and cost and time overruns (i.e., if any change in the complexity of a project will affect the project cost and time overruns). Finally, the regression models for predicting the cost and time overruns based on the complexity of a project were developed. The following table demonstrates descriptive statistics of the main attributes of fifty sample projects.

Table1: Descriptive Statistics of fifty sample projects

|  | Cost Overrun ($) | Time Overrun (days) |
|---|---|---|
| **Mean** | 19,943 | 18.64 |
| **Variance** | 203,594,899 | 167.87 |
| **Standard Deviation** | 14,268.67 | 12.96 |
| **Avg. % of cost/time overrun** | 5.7 | 7.3 |

These statistics show that the Changepoint projects have an average cost increase of 5.73%, with a maximum of 11.5% ($60,000) and a minimum of 0% (no cost overrun). Also, it has an average time overrun of 7.35%, with a maximum of 16.12% (50 days) and a minimum of 0% (no time overrun). Based on these statistics, the projects experienced a more time overrun compared to cost overrun. The confidence interval is calculated for the average cost and time overruns for all Changepoint's projects based on this sample data (assuming the confidence level is 95%):

$$\bar{x} \pm z_{\alpha/2} \frac{\sigma}{\sqrt{n}} \quad (1)$$

Since our sample size is larger than 30, the Normal Standard distribution is used. Confidence Interval for cost overrun:

> $\bar{x}$ = $19943    n = 50
> $\sigma$ = $14268.67
> [19943 – (1.96 * 14268.67 / 7.07),    19943 + (1.96 * 14268.67 / 7.07)]
> **CI for cost overrun = [15987.33, 23898.67]**

It means that we are 95% confident that the average of project cost overruns lies between $15,987.33 and $23,898.67.

> Confidence Interval for time overrun:
> $\bar{x}$ = 18.64 days    n = 50
> $\sigma$ = 12.96 days
> [18.64 – (1.96 * 12.96 / 7.07),    18.64 + (1.96 * 12.96 / 7.07)]
> **CI for time overrun = [15.05, 22.23]**

That means we are 95% confident that the average of projects time overruns lies between 15 and 22 days.

## VI. RESEARCH HYPOTHESES

The most commonly used regression method is linear regression analysis used to develop the predictive models. The models determine the impact of project complexity on the project cost and time overruns. Since the sample size is more than 30 (around 50 projects), we assume that data distribution is normal.

*Linear regression model for cost overrun:*
Project complexity is considered as independent variable and cost overrun as dependent variable. The regression equation is defined as below:

$Y_C = \beta_0 + \beta_1 X + e_C$
X = Project Complexity
$Y_C$ = Cost overrun
$\beta_0$ = Intercept
$\beta_1$ = Estimated regression coefficients of project complexity (slope)
$e_C$ = Error term

*Linear regression model for time overrun:*
Project complexity is considered as independent variable and time overrun (schedule slippage) as dependent variable. The regression equation is defined as below:

$Y_D = \beta_0 + \beta_1 X + e_D$
X = Project Complexity
$Y_D$ = Time delay
$\beta_0$ = Intercept
$\beta_1$ = Estimated regression coefficients of project complexity (slope)
$e_D$ = Error term

The following research hypotheses are defined as below:
First Hypothesis: There is a relationship between project complexity and cost overrun.

H₀ (Null Hypothesis):      $\beta_1 = 0$
H₁ (Alternative Hypothesis):      $\beta_1 \neq 0$

Second Hypothesis: There is a relationship between project complexity and time overrun.

H₀ (Null Hypothesis):      $\beta_1 = 0$
H₁ (Alternative Hypothesis):      $\beta_1 \neq 0$

## VII. LINEAR REGRESSION MODEL AND RESULTS

The data is plotted in the following scatter diagrams to display the relationship between cost and time overrun, and project complexity:

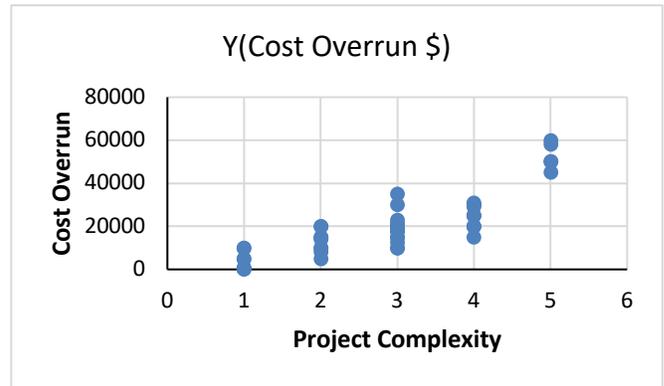

Figure1: Scatter plot of cost overrun ($) vs. project complexity

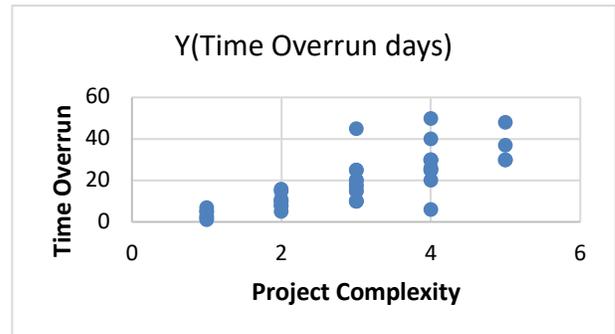

Figure2: Scatter plot of time overrun (days) vs. project complexity

The two regression models were examined using sample data through Excel. Based on the results (Table2 and Table3), the following regression equations were obtained.

The linear regression model for cost overrun (model1):
$Y_C = -8337.31 + 9751.83X + e_C$

In this model, the $\beta_1$ or slope is $9751.83, and the intercept is $-8337.31.

The linear regression model for time overrun (model2):
$Y_D = -6.20 + 8.56X + e_D$

In this model, the $\beta_1$ or slope is $8.56 days, and the intercept is $-6.20.

Table2: Coefficients and p-values for Cost Overrun model

|  | Coefficients | P-value |
|---|---|---|
| Intercept | -8337.30719 | 0.00338953 |
| Slope | 9751.830065 | 3.20971E-15 |

Table3: Coefficients and p-values for Time Overrun model

|  | Coefficients | P-value |
|---|---|---|
| Intercept | -6.197647059 | 0.024063719 |
| Slope | 8.564705882 | 1.54614E-13 |

## VIII. MODELS EXAMINATION, VALIDATION AND VERIFICATION

The output summary of regression analysis for two linear regression models is listed in Table4 and Table5:

Table4: Output summary of linear regression analysis for cost overrun (model1)

| Regression Statistics | |
|---|---|
| Multiple R | 0.853955728 |
| R Square | 0.729240385 |
| Adjusted R Square | 0.72359956 |
| Standard Error | 7501.581144 |
| Observations | 50 |

Since both model1 and model2 have one independent variable, we will look at the value of $R^2$ (coefficient of determination) and R (correlation coefficient) to determine our model is a good model. According to Table 4, the $R^2$ for the regression model1 is 0.73, suggesting that the variations in project complexity can explain 73% of the variation in the cost overrun. The regression model is satisfactory for prediction. Also, the correlation coefficient (R) for the regression model1 is +0.85, meaning a strong positive relationship between cost overrun and project complexity. This positive relationship can also be seen from the regression coefficient of the independent variable of project complexity (slope). This coefficient value is +9751.83 showing an increase in the project complexity leads to an increase in the cost overrun and vice versa.

Table5: Output summary of linear regression analysis for time overrun (Model2)

| Regression Statistics | |
|---|---|
| Multiple R | 0.825964706 |
| R Square | 0.682217695 |
| Adjusted R Square | 0.67559723 |
| Standard Error | 7.379482686 |
| Observations | 50 |

The regression model is satisfactory for prediction. According to Table5, the $R^2$ for the regression model1 is 0.68, suggesting that the variations in project complexity can explain 68% of the variation in the time overrun. Also, the correlation coefficient (R) for the regression model2 is +0.83, meaning a strong positive relationship between time overrun and project complexity. This positive relationship can also be seen from the regression coefficient for the independent variable of project complexity (slope). This coefficient value is +8.56 showing that an increase in the project complexity leads to an increase in project delay and vice versa.

According to Table 6, the significance of the F-test (the overall p for the model) is almost 0, less than any level of significance selected for statistical analysis ($\alpha$= 0.1, 0.05, or 0.01). In this case, there is strong evidence to reject the null hypothesis (H0) for the first hypothesis and accept the alternative hypothesis (H1). It means there is a relationship between cost overrun and project complexity. Based on the alternative hypothesis (H1), the test is two-tailed, and the calculated confidence interval for cost overrun slope (based on excel calculation) is [$8,027.36, $11,476.30]; we are confident by 95% that $\beta_1$ value ($9,753.83) lies between the confidence interval.

According to Table7, the significance of the F-test (the overall p for the model) is almost 0, less than any level of significance selected for statistical analysis ($\alpha$= 0.1, 0.05, or 0.01). In this case, there is strong evidence to reject the null hypothesis (H0) for the second hypothesis and accept the alternative hypothesis (H1). It means there is a relationship between time overrun and project complexity. Based on the alternative hypothesis (H1), the test is two-tailed, and the calculated confidence interval for time overrun slope (based on excel calculation) is [6.87 days, 10.26 days]; we are confident by 95% that $\beta_1$ value (8.56 days) lies between the confidence intervals. Further validation can predict the cost and time overruns for a project that is not part of our initial sample data and compare the result with actual cost/time overruns. The project data is listed in Table8.

Table 6: The analysis of variance for the regression Model 1

ANOVA

| | df | SS | MS | F | Significance F |
|---|---|---|---|---|---|
| Regression | 1 | 727501150 6 | 727501150 6 | 129.2790 2 | 3.20971E-15 |
| Residual | 48 | 270113854 4 | 56273719.6 6 | | |
| Total | 49 | 997615005 0 | | | |

Table7: The analysis of variance for the regression Model2

ANOVA

| | df | SS | MS | F | Significance F |
|---|---|---|---|---|---|
| Regression | 1 | 5611.595294 | 5611.595294 | 103.046799 | 1.54614E-13 |
| Residual | 48 | 2613.924706 | 54.45676471 | | |
| Total | 49 | 8225.52 | | | |

Table7: The analysis of variance for the regression Model2

Table8: Project A data which is used for further validation

| Project A | |
|---|---|
| Initial Budget($) | $365,000 |
| Complexity | 3 |
| Initial duration (days) | 225 days |
| Actual Cost($) | $386,500 |
| Actual Duration (days) | 245 days |
| Cost Overrun($) | $21,500 |
| Time Overrun(days) | 20 days |

Based on linear regression model1, we can predict cost overrun:

$Y_C$ = -8337.31 + 9751.83 * 3 = $20,918.18
$e_C$ = Y - $Y_C$ = 21,500 - 20,918.18 = $581.82

Based on linear regression model2, we can predict time overrun:

$Y_D = -6.20 + 8.56 * 3 = 19.48$ days
$e_D = Y - Y_D = 20 - 19.48 = 0.52$ day

The calculated error for cost overrun is $581.82 and for time overrun is 0.52 days which both are very small.

## IX. Summary and Conclusion

In summary, this study focuses on a statistical analysis of the impact of project complexity on cost overruns and schedule delays in Changepoint Company projects. The two hypotheses have been tested, one for checking the relationship between project complexity and cost overrun and the other for studying the relationship between project complexity and time overrun. The Null hypotheses were rejected in both tests, and there is enough evidence to accept the existence of the relationship between project complexity and cost/time overruns. Based on the descriptive analysis conducted on the data and regression modelling, there is a strong positive correlation between project complexity and cost/time overruns. When the project gets more complex, the chance of going over budget and having delays will be much higher and vice versa. Quantifying the strength of the relationship between complexity and cost/time overruns and developing and examining the predictive models have been done using fifty sample projects. A path forward for future studies would be to collect more empirical data from projects and obtain a larger sample size to examine the statistical models developed in this study thoroughly. This will also avoid any sampling bias and enhance the reliability of the outcomes. As a result of this study, the management team should focus more on the estimation and planning phase of Changepoint projects, particularly complex projects, to reduce cost and time overruns. Also, complex projects can be divided into smaller projects to be managed separately.

## References

[1]. Changepoint Corporation official website, Retrieved February 2016, from:
[2]. http://www.changepoint.com, Changepoint Press Release, October 2015, from: http://www.changepoint.com/getmedia/63655344-7ac6-4b40-a3ac-5443c875f28a/PR-Aug-21-2014-Changepoint-Daptiv
[3]. Ameh, O.J., &Osegbo, E.E. (2011), "Study of relationship between time overrun and productivity on construction sites", International Journal of Construction Supply Chain Management Volume 1 Number 1 2011, 56-67. 56.
[4]. Kaming, P. F., Olomolaiye, P. O., Holt, G. D., & Harris, F. C. (1997). Factors influencing construction time and cost overruns on high-rise projects in Indonesia. Construction Management & Economics, 15(1), 83-94.
[5]. Trigunarsyah, B. (2004). Constructability practices among construction contractors in Indonesia. Journal of construction engineering and management, 130(5), 656-669.
[6]. Elinwa, A. U., & Joshua, M. (2001). Time-overrun factors in Nigerian construction industry. Journal of construction engineering and management, 127(5), 419-425.
[7]. Bramble, B. B., & Callahan, M. T. (1987). Constmction Delay Claims.
[8]. Al-Momani, A. H. (2000). Construction delay: a quantitative analysis. International journal of project management, 18(1), 51-59.
[9]. Frimpong, Y., Oluwoye, J., & Crawford, L. (2003). Causes of delay and cost overruns in construction of groundwater projects in a developing countries; Ghana as a case study. International Journal of project management, 21(5), 321-326.
[10]. Ogunlana, S. O., Promkuntong, K., & Jearkjirm, V. (1996). Construction delays in a fast-growing economy: comparing Thailand with other economies. International Journal of project management, 14(1), 37-45.
[11]. Majid, M. A., & McCaffer, R. (1998). Factors of non-excusable delays that influence contractors' performance. Journal of management in engineering, 14(3), 42-49.
[12]. Akoa, B. B. (2011). Cost overruns and time delays in highway and bridge projects in developing countries: experiences from Cameroon. Michigan State University. Construction Management.
[13]. Chen, D., & Hartman, F. T. (2000). A neural network approach to risk assessment and contingency allocation. AACE International Transactions, RI7A.
[14]. Federle, M. O., & Pigneri, S. C. (1993). Predictive model of cost overruns. AACE International Transactions, L-7.
[15]. http://www.ijcscm.com/sites/default/files/issue/nid-6/eronini40_1324337978.pdf
[16]. Bertin Bella Akoa, 2011, "Cost Overruns and Time Delays in Highway and Bridge Projects in Developing Countries", Thesis, Michigan State University.
[17]. Jason J. Cook, Captain, USAF, AFIT/GEM/ENV/06M-02, "Estimating Required Contingency Funds for Construction Projects using Multiple Linear Regression Thesis", Department of the air force, Air University. Retrieved 2016.
[18]. "Cost Overrun", Wikipedia, Retrieved June 2016, from:
[19]. https://en.wikipedia.org/wiki/Cost_overrun
[20]. Mona Mahatha , "Software Development Risk Assessment: A multi-dimensional, Quantitative & Continuous Approach", PMI India National Conference 2015, PMI, India chapter, June 2015, from:
[21]. http://www.pmi.org.in/events/media/technical-paper/aframework-3e2cffd313b9b47.pdf
[22]. Tejale, D. S., Khandekar, S. D., & Patil, J. R. (2015). Analysis of construction project cost overrun by statistical method. International Journal, 3(5), 349-355.
[23]. Creedy, G. D., Skitmore, M., & Wong, J. K. (2010). Evaluation of risk factors leading to cost overrun in delivery of highway construction projects. Journal of construction engineering and management, 136(5), 528-537.
[24]. Larsen, J. K., Shen, G. Q., Lindhard, S. M., & Brunoe, T. D. (2016). Factors affecting schedule delay, cost overrun, and quality level in public construction projects. Journal of management in engineering, 32(1), 04015032.
[25]. Annamalaisami, C. D., & Kuppuswamy, A. (2021). Managing Cost Risks: Toward a Taxonomy of Cost Overrun Factors in Building Construction Projects. ASCE-ASME Journal of Risk and Uncertainty in Engineering Systems, Part A: Civil Engineering, 7(2), 04021021.
[26]. Amadi, A. I. (2021). Towards methodological adventure in cost overrun research: linking process and product. International Journal of Construction Management, 1-27.
[27]. babu Chandragiri, A., Jeelani, S. H., Akthar, S., & Lingeshwaran, N. (2021). A study and identification of the time and cost overrun in the construction project. Materials Today: Proceedings.

[28]. Pilger, J. D., Machado, E. L., de Assis Lawisch-Rodriguez, A., Zappe, A. L., & Rodriguez-Lopez, D. A. (2020). Environmental impacts and cost overrun derived from adjustments of a road construction project setting. Journal of Cleaner Production, 256, 120731.

[29]. hazal, M. M., & Hammad, A. (2020). Application of knowledge discovery in database (KDD) techniques in cost overrun of construction projects. International Journal of Construction Management, 1-15.

[30]. Hass, G., Simon, P., & Kashef, R. (2020, December). Business Applications for Current Developments in Big Data Clustering: An Overview. In 2020 IEEE International Conference on Industrial Engineering and Engineering Management (IEEM) (pp. 195-199). IEEE.

[31]. Ibrahim, A., Kashef, R., Li, M., Valencia, E., & Huang, E. (2020). Bitcoin network mechanics: Forecasting the btc closing price using vector auto-regression models based on endogenous and exogenous feature variables. Journal of Risk and Financial Management, 13(9), 189.

[32]. Kashef, R. (2021, April). Scattering-based Quality Measures. In 2021 IEEE International IOT, Electronics and Mechatronics Conference (IEMTRONICS) (pp. 1-8). IEEE

# APPENDIXES

**APPENDIX1:** The definition/the criteria defined for project complexity in Changepoint Company

| Complexity | Complexity dimension (initial duration/budget) |
|---|---|
| 1 | < 3 months, < $200,000 |
| 2 | 3 – 6 months, $200,000 – $300,000 |
| 3 | 6 – 9 months, $300,000 – $400,000 |
| 4 | 9 – 12 months, $400,000 – $500,000 |
| 5 | > 12 months, > $500,000 |